# What is the appropriate length of the publication period over which to assess research performance?[1]

Giovanni Abramo*

*Institute for Systems Analysis and Computer Science-National Research Council of Italy*

and

*Laboratory for Studies of Research and Technology Transfer*

*at University of Rome "Tor Vergata" – Italy*

ADDRESS: Dipartimento di Ingegneria dell'Impresa, Università degli Studi di Roma "Tor Vergata", Via del Politecnico 1, 00133 Roma - ITALY

tel. and fax +39 06 72597362, giovanni.abramo@uniroma2.it

Ciriaco Andrea D'Angelo

*Laboratory for Studies of Research and Technology Transfer*

*at University of Rome "Tor Vergata" – Italy*

ADDRESS: Dipartimento di Ingegneria dell'Impresa, Università degli Studi di Roma "Tor Vergata", Via del Politecnico 1, 00133 Roma – ITALY

tel. and fax +39 06 72597362, dangelo@dii.uniroma2.it

Tindaro Cicero

*Laboratory for Studies of Research and Technology Transfer*

*at University of Rome "Tor Vergata" – Italy*

ADDRESS: Dipartimento di Ingegneria dell'Impresa, Università degli Studi di Roma "Tor Vergata", Via del Politecnico 1, 00133 Roma – ITALY

tel. and fax +39 06 72595400, tindaro.cicero@uniroma2.it

* **Corresponding author**

# What is the appropriate length of the publication period over which to assess research performance?

**Abstract**
National research assessment exercises are conducted in different nations over varying periods. The choice of the publication period to be observed has to address often contrasting needs: it has to ensure the reliability of the results issuing from the evaluation, but also reach the achievement of frequent assessments. In this work we attempt to identify which is the most appropriate or optimal publication period to be observed. For this, we analyze the variation of individual researchers' productivity rankings with the length of the publication period within the period 2003-2008, by the over 30,000 Italian university scientists in the hard sciences. First we analyze the variation in rankings referring to pairs of contiguous and overlapping publication periods, and show that the variations reduce markedly with periods above three years. Then we will show the strong randomness of performance rankings over publication periods under three years. We conclude that the choice of a three year publication period would seem reliable, particularly for physics, chemistry, biology and medicine.

# 1. Introduction

The national research assessment exercises carried out in an ever growing number of nations (OECD, 2010), have different frequencies and cover periods of publication of different length. For example, in the United Kingdom, Research Assessment Exercises (the RAE series) have been held in 1986, 1989, 1992, 1996, 2001 and 2008, evidently assessing publication periods of different length. The next Research Excellence Framework (REF) refers to output produced over a 6-year publication period (1 January 2008 to 31 December 2013). In Australia, the Excellence in Research for Australia initiative (ERA, 2010) applied to research undertaken between 1 January 2003 and 31 December 2008, while the next ERA, 2012 will assess the period 1 January 2005 to 31 December 2010. In Italy, the first Triennial Evaluation Exercise (VTR) referred to the 3-year period 1 January 2001 to 31 December 2003, while the next Evaluation of Quality in Research (VQR) will cover the 7-year period 2004 to 2010. Aside from the Australian ERA, which has for the first time used an exclusively bibliometric approach for assessment of the hard sciences, all the above exercises are essentially based on the peer-review method, potentially informed by bibliometric indicators. This approach, aside from being very onerous, involves quite long execution times, generally two years. Given the great disparity in frequency, and in consequence also the publication periods assessed, the perception is that these choices are the result of organizational and budgetary restrictions, rather than the result of a technical decision arising from identification of the optimal length of observation period. This impression is supported by the observation that the next edition of the Australian ERA, based on the less onerous and more rapid bibliometric methodology, will now occur only two years after the preceding exercise. The design phase of any national exercise is very critical, since the results of the evaluation itself are sensitive to the project specifics (Geuna and Martin, 2003), such as the length of the publication period assessed, and in the case of use bibliometric indicators, the point in time for observation of citations. The current authors have previously inquired into the issue of the appropriate time length between citation counting and period of observation (Abramo et al., 2011a and 2011b). In this work we wish to identify, from a technical point of view, an appropriate length of publication period to be assessed in national research evaluation exercises.

Evaluation exercises are intended to accomplish various aims, as selected by the governments that adopt them: inform selective funding; stimulate better research performance; reduce information asymmetry between suppliers of new knowledge and their customers (students, companies, public administration, etc.); inform both national and regional research policies and research institutions' strategies; and last but not least, demonstrate that investment in research is effective and delivers public benefits. Whatever the objectives for which they are conducted, the greater is their frequency then the more effective they are in reaching their aim. Ideally one could conceive of assessment exercises conducted every year, based on the products of the preceding year. However, the production of new knowledge through research activity differs notably from other productive systems, both in manufacturing and services. Research projects have execution times that are generally longer, and notably different across disciplines and in fields within the same discipline. Since it is known that publication practices are sensitive to funding formulae (Bhattacharya and Newhouse, 2010; Butler, 2003), limiting the period of evaluation too much, i.e. to only one year, could have counterproductive effects on acceptance by researchers to take on the risks of diverting from consolidated scientific trajectories to explore new ones (Geuna and Martin, 2003). Further, both the time period from a paper's date of submission to a journal and its acceptance, and then from acceptance to actual publication date, are also highly variable within the same discipline, meaning that the shorter the publication period observed the greater the performance measures will be affected by a random component, external to the excellence of the researchers. The publication delays are more evident in the fields of mathematics and technical sciences (Luwel and Moed, 1998), in food research (Amat, 2008) and in econometrics (Trivedi, 1993).

Ultimately, measurements conducted over annual periods and with annual repetition, other than the risk of being counterproductive for scientific breakthroughs, would very likely be distorted, since they would be affected by this substantial random component. We view this component as particularly significant for evaluation exercises conducted at the level of individual scientists, a bit less for levels of analysis as they are increasingly aggregated (research teams, departments, universities). On the other hand, publication periods that are too long also present contraindications. Lengthening the period of observation, with consequent spacing out of frequency for exercises, may on the one hand ensure greater robustness in the rankings arising from the measure but on the other hand reduces the possibility of detecting variations in performance within the time period, and in consequence the effectiveness in achieving the evaluation objectives. The latest national research assessment exercises, conducted over periods of six or seven years with an added two years for execution, present a picture of an analyzed context that could very likely be changed at the moment for making decisions and undertaking interventions regarding the intended aims of the evaluation (selective funding, policy or strategy formulation, reduction of information asymmetry, etc.). The problem could be partially avoided by conducting more frequent evaluations, annual or biennial, over partially overlapping periods of $n$ years (for example triennial, quadrennial, five-year). However, again in this case, the longer the observation period $n$ then the less is the possibility of revealing detectable performance variations between one exercise and the next. For example, in an annually conducted five-year exercise, shifting the observation period by one year (in form 2003-2007; 2004-2008; 2005-2009,...), the production of the added year would have a relatively low weight with respect to that of the remaining years. The alternative is to conduct exercises over periods of $n$ years, with $n$ sufficiently large to cancel the random component or limit it as much as possible, but not so large as to hide variations in performance within the period. This evident trade-off between length of publication period evaluation and effectiveness of the evaluation with respect for the objectives for which it is conduced occurs regardless of the evaluation methodology adopted, characterizing both bibliometric exercises and those conducted by peer review.

The objective of this work is to evaluate the size of this tradeoff, investigating the sensitivity of individual researchers' productivity rankings to the length of the publication period, by means of bibliometric method, which is the only feasible approach at the experimental level. The literature does not seem to offer contributions useful for this purpose: scholars seem more focused on studies related to modeling citation age and mainly on comparing two basic approaches, namely the diachronous and the synchronous model (Glänzel, 2004; Burrell, 2002). The diachronous approach concerns measure of bibliometric indicators referring to a given period with variation in the time of observation of the citations, while synchronous studies fix the citation observation time but vary the length of the publication period. The analyses conducted consistently aim to compare the two approaches for purposes of improving the impact indicators of a single publication or journal. For the area of reference, the optimal publication period for national research assessment exercises, and for the specific intentions that inspire the work, once the moment of observing citations is fixed then the problem concerns the choice of the length of publication period over which to effect the analysis (synchronous approach). Thus the study will produce and compare different performance ranking lists corresponding to different scenarios of publication periods but with the moment of observing the citations always fixed. The object of the analysis is the research output in the 2003-2008 period by the research staff (more than 30,000 scientists) in the hard sciences at all Italian universities. We proceed by steps in our attempt to identify the most appropriate length of publication period. First we analyze the variation in rankings referring to pairs of contiguous and overlapping publication periods: beginning from 2008 (length = one year), we will repeatedly add one year, thus increasing the length of the publication period, until 2003 (length = six years, publication period 2003-2008). We will show that the variations reduce markedly, and always more, with publications periods beginning from a three year period and greater. At this point we will then concentrate on measurement of variations in productivity between contiguous triennia: to understand if and how much such variations are truly representative of structural trend, we will again conduct the analysis for single years and then for biennia. This final analysis will show the strong randomness of performance rankings over publication periods under three years.

The following section of the work describes the methodology applied to the analysis and the dataset used. Section 3 shows the elaborations and the results concerning the variation of individual researchers' productivity rankings with the length of the publication period. The final section comments on the main findings and their implications.

## 2. Methodology and dataset

To compare productivity of individual researchers we consider the value of output, i.e. the impact, of their research activities in a given period of time, whose length varies over the period 2003 to 2008. As proxy of overall output per researcher in the hard sciences, we consider publications (articles, article reviews and conference proceedings) indexed in Web of Science (WoS). As proxy of the value of output we adopt the number of citations for the researchers' publications at 30/06/2009. Because the intensity of publications varies by field, we compare researchers within the same field and rank them in percentile scale (Abramo and D'Angelo, 2011). In the Italian university system all research personnel is classified in one and only one field. In the hard sciences, there are 205 such fields (named scientific disciplinary sectors, SDSs[2]), grouped into nine disciplines (named university disciplinary areas, UDAs[3]). We note that it is not unlikely that researchers belonging to a particular scientific field may also publish outside that field. Because citation behavior varies across fields (Abramo and D'Angelo, 2007; Moed, 2005) we standardize the citations for each publication, with respect to the median[4] of the distribution of citations for all the Italian publications of the same year and the same WoS subject category[5]. When measuring labor productivity, should there be differences in the production factors available to each scientist, one should normalize by them. Unfortunately relevant data are not available. The productivity of a single researcher, named Fractional Scientific Strength (FSS)[6], is given by:

$$FSS = \sum_{i=1}^{N} \frac{c_{ij}}{Me_j} * \frac{1}{s_i}$$

---

[2] The complete list is accessible on http://attiministeriali.miur.it/UserFiles/115.htm. Last accessed on March 12, 2012.

Where:
$c_{ij}$ = citations received by publication i, falling in subject category *j*;
$Me_j$ = median of the distribution[7] of citations received for all Italian publications of the same year and subject category *j*;
$s_i$ = co-authors[8] of publication *i*
N = number of publications of the researcher in the period of observation.

We elaborate researchers' FSS ranking lists for each SDS, considering different lengths of publication periods from 2003 to 2008.

Data on research staff of each university and their SDS classification is extracted from the database on Italian university personnel, maintained by the Ministry for Education, Universities and Research[9]. The bibliometric dataset used to measure FSS is extracted from the Italian Observatory of Public Research (ORP)[10], a database developed and maintained by the authors and derived under license from the Thomson Reuters WoS. Beginning from the raw data of the WoS, and applying a complex algorithm for reconciliation of the author's affiliation and disambugation of the true identity of the authors, each publication is attributed to the university scientist or scientists that produced it (D'Angelo et al., 2010).

---

[3] Mathematics and computer sciences; physics; chemistry; earth sciences; biology; medicine; agricultural and veterinary sciences; civil engineering; industrial and information engineering.

[4] As frequently observed in literature (Lundberg, 2007), standardization of citations with respect to median value rather than to the average is justified by the fact that distribution of citations is highly skewed in almost all disciplines.

[5] The subject category of a publication corresponds to that of the journal where it is published. For publications in multidisciplinary journals the median is calculated as a weighted average of the standardized values for each subject category.

[6] This indicator is similar to the "total field normalized citation score" of the Karolinska Institute (Rehn et al., 2007). The difference is that we standardize by the Italian median rather than the world average. Moreover, we consider fractional counting of citations based on co-authorship.

[7] Publications without citations are excluded from calculation of the median.

[8] For life sciences, different weights are given to each co-author according to his/her position in the list and the character of the co-authorship (intra-mural or extra-mural). If first and last authors belong to the same university, 40% of citations are attributed to each of them; the remaining 20% are divided among all other authors. If the first two and last two authors belong to different universities, 30% of citations are attributed to first and last authors; 15% of citations are attributed to second and last author but one; the remaining 10% are divided among all others.

[9] http://cercauniversita.cineca.it/php5/docenti/cerca.php. Last accessed on March 12, 2012.

[10] www.orp.researchvalue.it/ Last accessed on March 12, 2012.

To ensure the representativity of publications as proxy of the research output, the fields of observation are limited to those SDSs where at least 50% of researchers produced at least one publication in the period 2003-2008. Furthermore, we excluded those SDSs with fewer than 10 members. In total we finally considered 181 SDSs. The dataset is thus composed of 196,996 publications authored by a total of 30,611 academic scientists in continuous faculty role over the six years under observation: Table 1 shows their distribution in the SDSs of the nine UDAs considered.

*Table 1: Number of universities, research staff, publications and SDSs per UDA; 2003-2008 data*

| UDA | Universities | SDSs | Research staff | Publications | Authorship |
|---|---|---|---|---|---|
| Mathematics and computer sciences | 61 | 9 | 2,841 | 16,297 | 19,950 |
| Physics | 60 | 8 | 2,284 | 28,033 | 63,451 |
| Chemistry | 58 | 11 | 2,827 | 29,101 | 55,767 |
| Earth sciences | 48 | 12 | 1,100 | 5,422 | 7,369 |
| Biology | 63 | 19 | 4,474 | 33,101 | 53,367 |
| Medicine | 55 | 46 | 9,333 | 58,929 | 115,189 |
| Agricultural and veterinary sciences | 46 | 28 | 2,446 | 11,786 | 20,628 |
| Civil engineering | 47 | 7 | 1,126 | 4,939 | 6,287 |
| Industrial and information engineering | 63 | 41 | 4,180 | 37,082 | 53,094 |
| Total | 72 | 181 | 30,611 | 196,996* | 395,102 |

*\* Totals of data from the columns differ from the values shown in the last line due to the effect of multiple counts for publications co-authored by scientists belonging to SDSs from different UDAs*

## 3. Results and analysis

With the aim of helping decision makers to identify the appropriate compromise between greater frequency of execution for evaluation exercise and greater robustness of performance rankings, we simulate several bibliometric assessment scenarios, varying the length of the publication period, but fixing the citation observation time on 30/06/2009.

### 3.1 Performance differences when increasing the length of the publication period

We begin by analyzing variation of the performance rankings for pairs of contiguous and overlapping publication periods, increasing the length of publication period by one year, starting from 2008 (length = one year) up to 2003 (length = six years). The reference context is that of a decision maker who, on 30/06/2009, must choose which publication period (and thus which set of publications) to conduct an assessment. The first option is to consider a single year, 2008, the year just concluded; the second option would be the biennium 2007-2008, and so on, going backwards and step by step adding another year.

For each scenario we compute the FSS ranking lists of the researchers of each SDS. Subsequently we classify researchers into four classes, according to their ranking, assigning values of 4, 3, 2 and 1, corresponding to the first, second, third and fourth quartiles for the productivity distribution in each SDS. A value of 0 is assigned to researchers that are unproductive (FSS=0), either because they did not publish or did not receive any citations.

As an example, Table 2 provides the ranking list for the six scenarios considered for researchers in ING-IND/05-Aerospace systems, an SDS where there are only 26 researchers in stable faculty role over the six years considered. The results show that the first three of the list are at the top of the SDS ranking in all six of the scenarios considered. Another six researchers (Researcher_ID 6, 7, 8, 9, 10 and 12) are inactive in both 2007 and 2008 but three of them place in the third quartile and three in the second performance quartile if evaluated over the full six years of the publication period. At the bottom of the group are 11 scientists (ID 16-26) who are constantly inactive in all six years considered. These researchers are not considered in further analysis, since their performance does not change in function of the publication period considered.

*Table 2: Productivity rankings (quartiles: 4 = top) of Italian academic researchers in the SDS 'Aerospace Systems'*

| Researcher_ID | 2008 | 2007-2008 | 2006-2008 | 2005-2008 | 2004-2008 | 2003-2008 |
|---|---|---|---|---|---|---|
| 1 | 4 | 4 | 4 | 4 | 4 | 4 |
| 2 | 4 | 4 | 4 | 4 | 4 | 4 |
| 3 | 4 | 4 | 4 | 4 | 4 | 4 |
| 4 | 4 | 3 | 4 | 4 | 4 | 4 |
| 5 | 0 | 4 | 2 | 2 | 3 | 3 |
| 6 | 0 | 0 | 3 | 3 | 3 | 3 |
| 7 | 0 | 0 | 0 | 3 | 3 | 3 |
| 8 | 0 | 0 | 0 | 0 | 0 | 3 |
| 9 | 0 | 0 | 3 | 3 | 2 | 2 |
| 10 | 0 | 0 | 3 | 3 | 2 | 2 |
| 11 | 0 | 3 | 2 | 3 | 3 | 2 |
| 12 | 0 | 0 | 3 | 2 | 2 | 2 |
| 13 | 0 | 4 | 3 | 2 | 2 | 1 |
| 14 | 0 | 0 | 2 | 2 | 1 | 1 |
| 15 | 0 | 0 | 0 | 0 | 1 | 1 |
| 16-26 | 0 | 0 | 0 | 0 | 0 | 0 |

The analysis was repeated for all 181 SDSs considered: Figure 1 shows the average value of differences by quartiles (by UDA), between adjacent scenarios. In the comparison between the 2008 and the 2007-2008 rankings lists, this value varies between a minimum of 0.7 for chemistry and a maximum of 1.5 for civil engineering. We observe a significant discontinuity of the curve corresponding to the comparison between scenarios 2 and 3 (2007-2008 vs. 2006-2008). After this point the variations of slope reduce and the curve seems to linearize: with the exception of Agricultural and veterinary sciences, the UDA where discontinuity is more evident corresponding to the comparison between scenarios 4 and 5, and civil engineering, where the curve once again rises at the comparison between the last two scenarios. In reality, the plot shows a certain clusterization of the UDAs: industrial and information engineering, mathematics and computer sciences, earth and space sciences, in addition to previously noted agricultural and veterinary sciences and civil engineering, show a less obvious "elbow" at the second point on the curve compared to a second cluster composed of biology, chemistry, medicine and physics, where the change in slope is more clear. The results just described are in line with the results of a correlation analysis conducted between rankings from the different simulation scenarios. Table 3 presents the Spearman correlation values between pairs of adjacent scenarios. Comparing the rankings constructed over a publication period of two years (scenario 2, 2007-2008) and over 3 years (scenario 3, 2006-2008), we note extremely high correlation values, with a maximum in physics (+0.882) and a minimum in agricultural and veterinary sciences (+0.767). In the next comparison, between the rankings from scenarios 3 and 4, the correlation values are all over 0.8.

In conclusion, it would seem that the results of a potential bibliometric assessment would already be relatively "stabilized" with a publication period of three years: adding further years would certainly refine the rankings but with ever more marginal incremental effects. This seems most evident in medicine, physics, biology and chemistry.

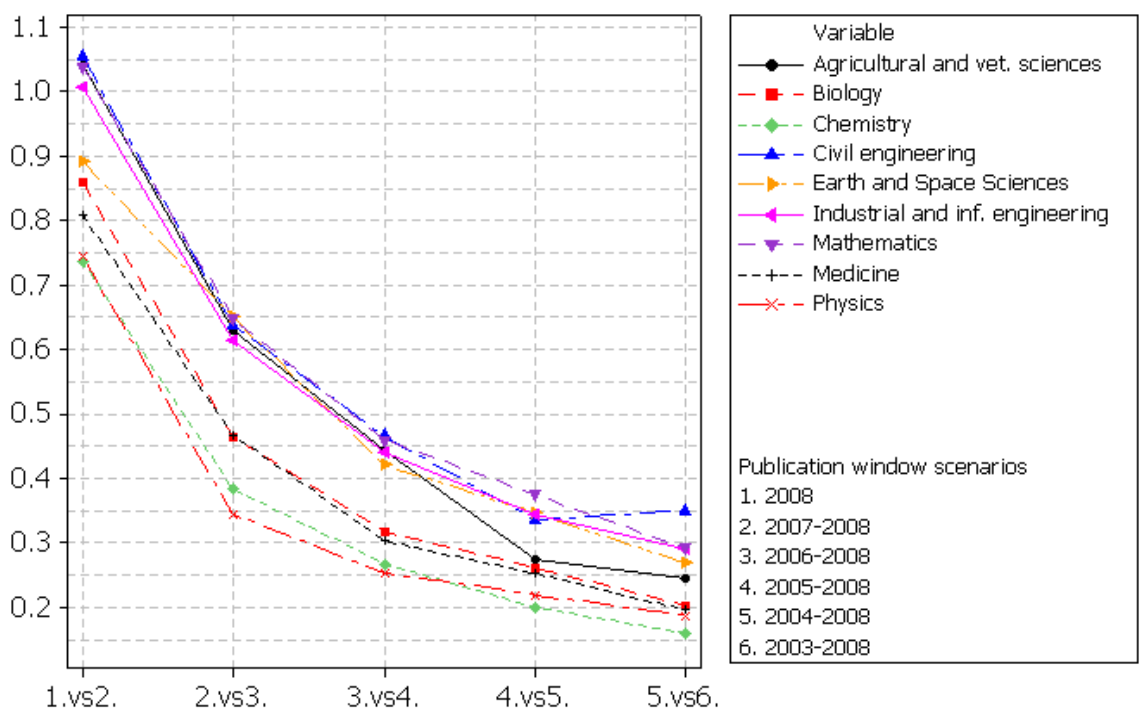

*Figure 1: Average quartile differences comparing six contiguous scenarios of publication period, by UDA*

*Table 3: Spearman correlations between quartile rankings of contiguous scenarios by UDA*

| UDA | 1 vs. 2 | 2 vs. 3 | 3 vs. 4 | 4 vs. 5 | 5 vs. 6 |
|---|---|---|---|---|---|
| Mathematics and computer sciences | +0.598 | +0.787 | +0.833 | +0.861 | +0.891 |
| Physics | +0.745 | +0.882 | +0.914 | +0.919 | +0.927 |
| Chemistry | +0.755 | +0.864 | +0.904 | +0.925 | +0.937 |
| Earth sciences | +0.691 | +0.774 | +0.840 | +0.876 | +0.889 |
| Biology | +0.701 | +0.837 | +0.888 | +0.900 | +0.921 |
| Medicine | +0.716 | +0.841 | +0.893 | +0.906 | +0.928 |
| Agricultural and veterinary sciences | +0.613 | +0.767 | +0.836 | +0.894 | +0.904 |
| Civil engineering | +0.584 | +0.776 | +0.839 | +0.877 | +0.852 |
| Industrial and information engineering | +0.631 | +0.782 | +0.833 | +0.867 | +0.886 |

## 3.2 Longitudinal analysis

As seen in the preceding section, a publication period of three years seems sufficient to guarantee a certain robustness in rankings resulting from bibliometric research performance evaluation. But what are the contraindications against a greater publication period? The lengthening of the publication period for the evaluation exercises, and the resulting lesser frequency of these, on the one hand could offer greater robustness in the rankings arising from the assessment, but on the other would reduce possibilities to detect performance variations appearing within the period, with all the consequences concerning the objectives of the research assessments. In this section we show the differences in an evaluation conducted over six years, compared to one conducted for two consecutive triennia. For this, we imagine a first evaluation conducted over the entire 2003-2008 publication period and a second conducted over its two constituent triennia: 2003-2005 and 2006-2008. We focus attention on a specific subpopulation of researchers, the 'top scientists', or those who register performance in the highest quartile[11]. What we measure here is the variation in performance of this subpopulation between the two triennia compared to that emerging from the overall evaluation over the six years: we use "increasing" to define performance of those who classify as top scientist in the second triennium, coming from a lower quartile in the first triennium. Vice versa, "decreasing" defines performance of researchers classified as top in the first triennium but not in the second. Table 4 presents the data on the percentage of researchers with increasing or decreasing performance, which would not be observed in a six year evaluation. In particular, column 2 shows the percentage of researchers that present a worsening in personal performance between the two contiguous triennia, but are classified as top scientists in the 6 year evaluation.

*Table 4: Percentage of researchers with increasing or decreasing performance between two triennia*

| UDA | Decreasing performance | Increasing performance | Total |
|---|---|---|---|
| Mathematics and computer sciences | 6.9% | 6.9% | 13.8% |
| Physics | 5.4% | 5.1% | 10.5% |
| Chemistry | 4.3% | 4.3% | 8.6% |
| Earth sciences | 8.0% | 8.0% | 16.0% |
| Biology | 4.7% | 4.7% | 9.4% |
| Medicine | 4.9% | 4.9% | 9.8% |
| Agricultural and veterinary sciences | 6.4% | 6.6% | 13.0% |
| Civil engineering | 6.1% | 6.1% | 12.2% |
| Industrial and information engineering | 7.1% | 7.1% | 14.2% |

In earth sciences, an evaluation conducted over 2003-2008 would ignore the fact that 8.0% of researchers who qualified as top scientist actually had important worsening in their performance. Column three, meanwhile, shows the percentage of scientists that move into top category from the first to second triennium, but are classified as such over the full six year evaluation. In total (last column of the table) we observe that in five UDAs (earth sciences, industrial engineering, civil engineering, mathematics, agricultural and veterinary sciences), a six year evaluation is not able to detect very important variations in performance that occur (for bad or good) in the two intermediate triennia, for over 12% of the total researchers.

---

[11] This is generally the category taken as reference in defining the specifics of incentive systems.

### 3.3 The random variation of ranks

In the preceding section it emerged that in lengthening the publication period (for example to six years), it is not possible to perceive structural performance variations that occur over time (for example between adjacent triennia). However there still remains a doubt about what part of the variations observed between these triennia could simply be due to randomness. In other words, we ask whether a publication period of three years would be sufficient to eliminate the effect of the random component on value of research productivity. To answer, we evaluate the performance of three researches from different SDSs and UDAs, representing three typical cases: increasing, decreasing and constant performance. We compare the results from an evaluation conducted for single years, for biennia and then for triennia. In Figure 2, for each publication period with equal length (1, 2 and 3 years), we show the historic series of performance and the regression (dashed line) that would represent the structural component of change, or the trend in actual performance. In the evaluation conducted year by year, Researcher#1, in BIO/17 (histology), places in the second quartile in three years (2004, 2006 e 2007), but in others has a higher performance (third quartile in 2003 and fourth in 2005), and in the last year results as inactive. Performance by Researcher#2 also swings quite strongly, ranging from the head of the rankings in his SDS (CHIM/08-pharpmaceutical chemistry) from 2006 to 2008, while in the first biennium (2003-2004) he was fixed at the second quartile and even dropped to last in 2005. Researcher#3 is an example of rapid and cyclical performance, occupying the top quartile of his SDS (MED/04-general pathology) in 2003 and 2007, and never remaining in the same quartile for consecutive years.

If we instead consider a publication period of two years, the oscillations are less relevant, however any possibility of identifying a true trend is still not at all a given. In fact, Researcher#1 progresses from quartile 3 in the first biennium to top quartile in the second, but then descends again to second in the last biennium; Researcher#2 places in the last quartile in the first biennium, to then bounce to first in the two successive biennia; Researcher#3, however, starts from quartile 3 in the first biennium, drops to quartile 2 in the next biennium, then rises to the top at the last.

The three final diagrams, concerning triennial evaluations, permit an appreciation of a substantial change similar to the dashed trend line in the diagrams for annual and biennial evaluation, and without the disturbance by evident random oscillations seen for the shorter periods.

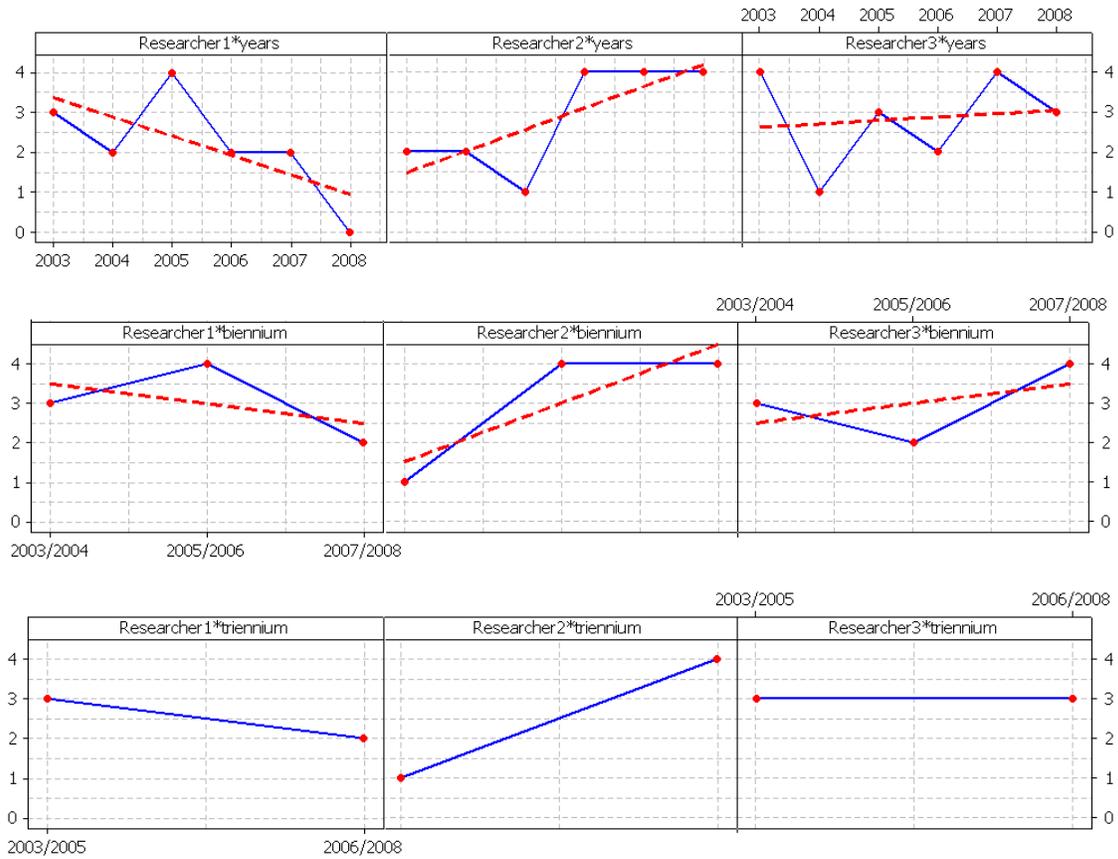

*Figure 2: Time series of performance (quartiles) for three researchers for publication periods respectively of one, two and three years, over the period 2003 to 2008*

Summing up, we have $\Delta_{oss} = \Delta_{strutt} + \varepsilon$, where:
$\Delta_{obs}$ : Observed performance variation
$\Delta_{real}$ : Real performance variation
ε = random component of variation

By definition, the random component (thus the value of ε) lessens with increase in the publication period: but we cannot be certain that a publication period of three years would be sufficient to render it negligible. To verify, we extend the analysis of the previous example to all researchers of the dataset and compare the variations detected between contiguous publication periods: annual, biennial, and triennial. Table 5 shows the results for each UDA:

- the second column shows the average variation of ($\Delta_1$) for the comparison between the six annual rankings (2003 vs. 2004, 2004 vs. 2005, … , 2007 vs. 2008).
- column three shows the average variation of quartile ($\Delta_2$) for the comparison between the three biennial rankings (2003-2004 vs. 2005-2006, 2005-2006 vs. 2007-2008)
- column three shows the variation of quartile ($\Delta_3$) for the comparison between the two triennial rankings (2003-2005 vs. 2006-2008).

In keeping with the observations in the example of the three researchers, the variations between adjacent years are strongly affected by chance: the data in column 2 show that this is somewhat evident in all the UDAs, though with significant differentiations: in mathematics and computer sciences, civil engineering and earth sciences, the average variation of quartile for annual evaluation is greater than 1.5; in physics and chemistry it is slightly less than one; in all other UDAS it takes intermediate values. Passing to the biennial publication period, the average variations of performance ($\Delta_2$) reduces in significant manner, except for civil engineering. In the other four UDAs (mathematics and computer sciences, earth sciences, agricultural and veterinary sciences, industrial and information engineering), the average variation of quartile remains over one. We can still argue that such values increase the weight of the structural component of performance variation at the expense of the random component. In the comparison between triennia, the variation ($\Delta_3$) decreases further, however still remaining greater than one in five UDAs and not particularly far from one in the remaining four: we can thus imagine that the variation detected would be stabilizing around the trend value ($\Delta_{obs} \cong \Delta_{real}$) and that the random component ($\varepsilon$) now has only marginal effect on the entity of shifts that are seen.

*Table 5: Average difference of performance quartile relative to contiguous publication periods of one, two and three years, in the period 2003-2008*

| UDA | $\Delta_1$ | $\Delta_2$ | $\Delta_3$ |
|---|---|---|---|
| Mathematics and computer sciences | 1.522 | 1.351 | 1.231 |
| Physics | 0.967 | 0.762 | 0.734 |
| Chemistry | 0.994 | 0.739 | 0.670 |
| Earth sciences | 1.508 | 1.371 | 1.248 |
| Biology | 1.242 | 0.960 | 0.854 |
| Medicine | 1.146 | 0.999 | 0.915 |
| Agricultural and veterinary sciences | 1.457 | 1.293 | 1.142 |
| Civil engineering | 1.544 | 1.532 | 1.452 |
| Industrial and information engineering | 1.392 | 1.237 | 1.160 |

We further note that the UDAs that show $\Delta_3$ less than one (physics, chemistry, biology, medicine) are the same as the ones in the cluster indicated in Figure 1, confirming that the correlation between length of publication period and reliability of results of an evaluation varies on the basis of the specificity of the field considered, and in particular depends on the intensity of publication and citation of the field itself.

**4. Conclusions**

A range of nations have now long implemented national research assessment exercises as stable practice, but with different frequencies, covering publication periods of different lengths. This variety of solutions indicates that the choices made do not respond to rigorous criteria for optimization of the tradeoff between robustness and functionality of rankings, but rather to the presence of budgetary and/or organizational restrictions that differ from nation to nation. This is confirmed by the fact that, while almost all of these exercises are conducted by the peer review approach, typically characterized by elevated costs and times for execution, the recent choice of a bibliometric approach for the Australian ERA has permitted the scheduling of the next evaluation exercise after a gap of only two years from the preceding one.

Beginning from these observations, in this work we have tried to identify an appropriate length for the publication period of a hypothetical national research evaluation exercise. In selecting the word "appropriate", we wish to qualify this as a choice free of time or financial budget restrictions, and exclusively the fruit of optimization of the noted tradeoff.

In this regard, independent of the objectives for which such assessments are realized (for example, stimulating better research performance), the greater is their frequency then the greater the effectiveness. However, greater frequency corresponds to lesser length of publication period, and restricting the performance observation period too much could have various counter-indications. It could determine relevant conditioning in choices by the scientists, particularly discouraging them from undertaking new paths of research along unexplored scientific trajectories. There is also a very relevant risk that measurement conducted over very short publication periods could be affected by a consistent component of randomness. The times for maturing results from research activity, already long when compared to other production systems, also vary across fields and are often conditioned by factors that are not correlated to the skill of those who produced them.

Until now, this evident tradeoff has not received significant investigation in the literature. For this, the authors have attempted to investigate the sensitivity of individual researchers' productivity rankings to the length of the publication period, done by means of the bibliometric method, which is the only feasible approach at the experimental level.

The analysis conducted revealed that the results of bibliometric assessment would already seem reliable with a publication period of three years, above all in medicine, physics, biology and chemistry. Further increases in the length of the publication period would certainly refine the rankings, however the incremental effects are always less significant. Further, it becomes step by step more difficult to detect very important performance variations that occur over the course of the publication period, for very consistent shares of the population, a situation detrimental to the actual functionality for assessment objectives, which are first of all to stimulate continuously better performance.

A doubt that a publication period of three years might not be sufficient to reasonably reduce the effects of the random component of value of productivity was relieved by an analysis of the variations in productivity between contiguous triennia. The variations observed seem truly representative of structural trends for change, bad or good, in the performance of individual researchers. Again, this appears more evident in medicine, physics, biology and chemistry. In the future the decision maker at the political or institutional level, responsible for the design of a research assessment exercise, can not only take account of budgetary and organizational considerations, but also the results of the present study.

# References

Abramo, G., Cicero, T., D'Angelo, C.A. (2011a). Assessing the varying level of impact measurement accuracy as a function of the citation window length. *Journal of Informetrics,* 5(4), 659-667.

Abramo, G., Cicero, T., D'Angelo, C.A. (2012) A sensitivity analysis of researchers' productivity rankings to citation window length. *Journal of* Informetrics, 6(2), 192-201.

Abramo, G., D'Angelo, C.A. (2011). National-scale research performance assessment at the individual level. *Scientometrics*, 86(2), 347-364.

Abramo, G., D'Angelo, C.A., (2007). Measuring science: irresistible temptations, easy shortcuts and dangerous consequences. *Current Science*, 93(6), 762-766.

Amat, C.B. (2008). Editorial and publication delay of papers submitted to 14 selected Food Research journals. Influence of online posting. *Scientometrics*, 74(3), 379-389.

Bhattacharya, A., Newhouse, H. (2010). Allocative Efficiency and an Incentive Scheme for Research. *Discussion Papers 10/02*, Department of Economics, University of York.

Burrell, Q.L. (2002). Modeling citation age data: Simple graphical methods from reliability theory. *Scientometrics*, 55, 273-285.

Butler, L. (2003). Modifying publication practices in response to funding formulas. *Research Evaluation*, 17(1), 39-46.

D'Angelo, C.A., Giuffrida, C., Abramo, G. (2011). A heuristic approach to author name disambiguation in large-scale bibliometric databases, *Journal of the American Society for Information Science and Technology*, 62(2), 257–269.

ERA (2010). The Excellence in Research for Australia (ERA) Initiative, http://www.arc.gov.au/era/ last accessed on December 16, 2011.

Geuna, A., Martin, B.R. (2003). University research evaluation and funding: an international comparison. *Minerva*, 41(4), 277–304.

Glänzel, W. (2004). Towards a model for diachronous and synchronous citation analyses. *Scientometrics*, 60(3), 511-522.

Lundberg, J. (2007). Lifting the crown – citation z-score. *Journal of Informetrics,* 1(2), 145-154.

Luwel, M., Moed, H.F. (1998). Publication delays in the science field and their relationship to the ageing of scientific literature. *Scientometrics,* 41(1-2), 29-40.

Moed, H.F. (2005). *Citation Analysis in Research Evaluation*. Springer, ISBN: 978-1-4020-3713-9

OECD (2010). Performance-based funding for public research in tertiary education institutions, *Workshop proceedings OECD 2010,* 187, ISBN 978-92-64-09461-1.

Rehn, C., Kronman, U., Wadsko, D. (2007). Bibliometric indicators definitions and usage at Karolinska Institutet. *Karolinska Institutet University Library*. Last accessed on March 8, 2012 from http://kib.ki.se/sites/kib.ki.se/files/Bibliometric_indicators_definitions_1.0.pdf

Trivedi, P.K. (1993). An analysis of publication lags in econometrics. *Journal of Applied Econometrics*, 8(1), 93-100.